\documentclass[conference]{IEEEtran}

\usepackage[mathcal]{eucal}
\usepackage{mathrsfs}
\usepackage{bm}

\usepackage{cite}

\ifCLASSINFOpdf
\usepackage[pdftex]{graphicx}
\else
  \usepackage[dvips]{graphicx}
\fi
\usepackage[cmex10]{amsmath}
\usepackage{amssymb}
\ifCLASSOPTIONcompsoc
  \usepackage[caption=false,font=normalsize,labelfont=sf,textfont=sf]{subfig}
\else
  \usepackage[caption=false,font=footnotesize]{subfig}
\fi

\usepackage{bm}
\usepackage{amssymb}

\usepackage{epstopdf}

\usepackage{color}

\def\BibTeX{{\rm B\kern-.05em{\sc i\kern-.025em b}\kern-.08em
    T\kern-.1667em\lower.7ex\hbox{E}\kern-.125emX}}
\begin{document}

\title{Federated AirNet: Hybrid Digital-Analog Neural Network Transmission for Federated Learning}


\author{
\IEEEauthorblockN{
Takuya Fujihashi\IEEEauthorrefmark{1},
Toshiaki Koike-Akino\IEEEauthorrefmark{2},
Takashi Watanabe\IEEEauthorrefmark{1}
}

\IEEEauthorblockA{
\IEEEauthorrefmark{1}Graduate School of Information Science and Technology, Osaka University, Japan\\
}
\IEEEauthorblockA{
\IEEEauthorrefmark{2}Mitsubishi Electric Research Laboratories (MERL), 201 Broadway, Cambridge, MA 02139, USA
}
}

\maketitle

\begin{abstract}
A key issue in federated learning over wireless channels is how to exchange a large number of the model parameters via time-varying channels. 
Two types of solutions based on digital and analog schemes are used typically. 
The digital-based solution takes quantization and entropy coding for compression, whereas transmissions via wireless channels may cause catastrophic errors owing to the all-or-nothing behavior in entropy coding.
The analog-based solutions such as AirNet and AirComp use analog modulation for the parameter transmissions. 
However, such an analog scheme often causes significant distortion due to the source signal's large power without compression gain.
This paper proposes a novel hybrid digital-analog transmission---Federated AirNet---for the model parameter transmissions in federated learning. 
The Federated AirNet integrates low-rate digital coding and energy-compact analog modulation. 
The digital coding offers the baseline of the model parameters and compacts the source signal power. 
In addition, the residual parameters, which are obtained from the original and encoded model parameters, are analog-modulated to enhance the baseline according to the instantaneous wireless channel quality. 
We show that the proposed Federated AirNet yields better image classification accuracy compared with the digital-based and analog-based solutions over a wide range of wireless channel signal-to-noise ratios (SNRs).
\end{abstract}


\section{Introduction}
\begin{figure}[t]
  \begin{center}
   \includegraphics[width=\hsize]{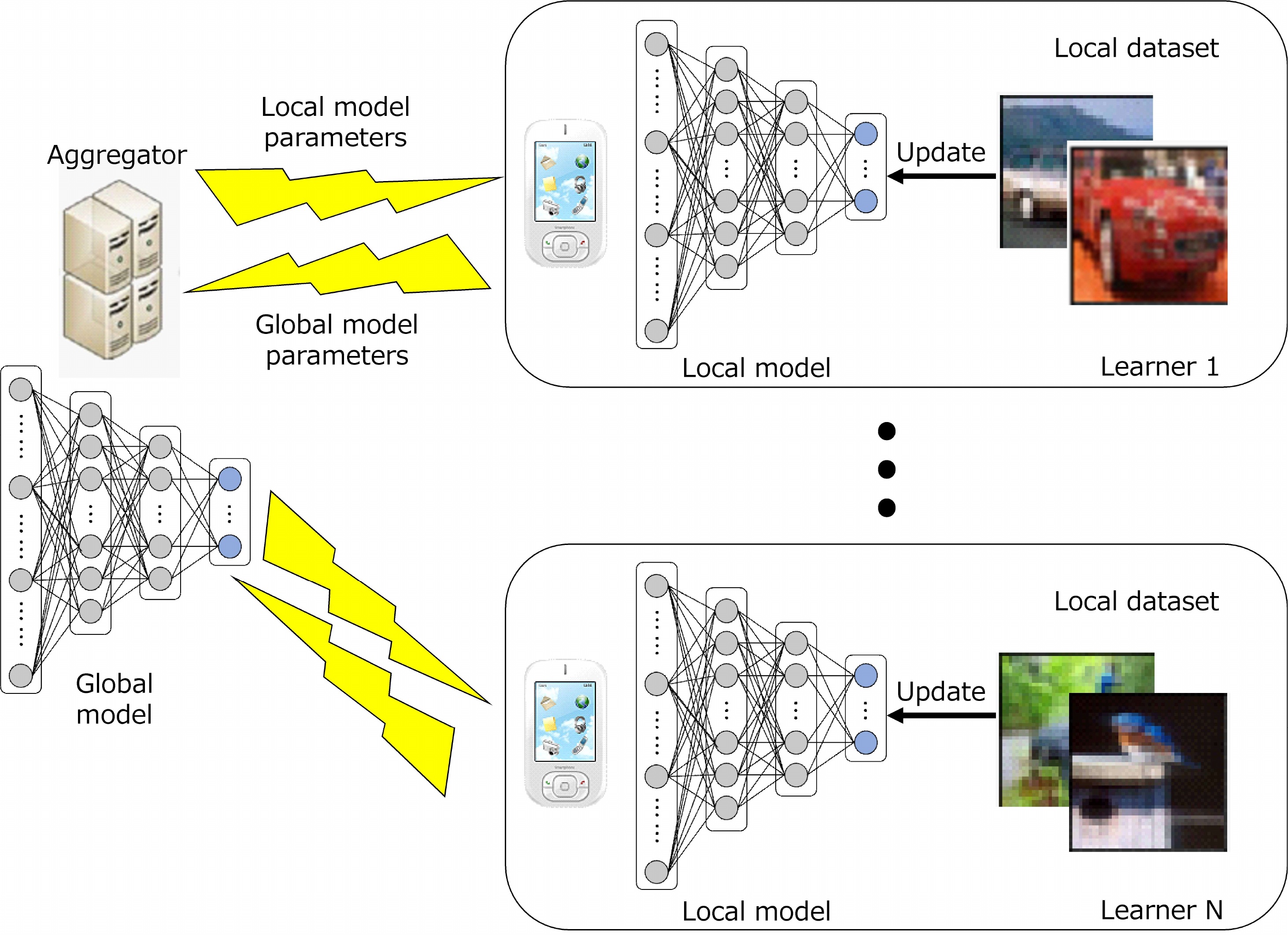}
   \caption{Overview of the federated learning with $N$ distributed learners in the federation and one aggregator over wireless networks.}
   \label{fig:FL}
  \end{center}
\end{figure}

Although the deep neural networks~(DNN) can solve various tasks using a large dataset, handling such a dataset increases costs and power consumption~\cite{strubell2019energy}, especially in wireless edge devices due to the limited power and bandwidth. 
Federated learning (FL)~\cite{bib:FL1,bib:FL2} is a decentralized learning approach to solve the issues mentioned above. 
In FL, model training is carried out over a federation of distributed learners and an aggregator. 
Fig.~\ref{fig:FL} shows one typical FL example over wireless networks.
Each learner in the federation uses only locally available data for training. 
After training its local model, each learner sends the updated parameters to the aggregator instead of the local data. 
The aggregator updates the global model based on the received parameters from the distributed learners, and the updated global model parameters are regularly shared back to the distributed learners. 
Even though each learner only has locally available data, it can benefit from the shared global model without explicitly accessing other learners' data.

One critical issue in FL over wireless networks is how to exchange a large number of model parameters between the aggregator and learners in the federation. 
The simplest solution is to quantize and binarize the parameters using a digital source coding. 
DeepCABAC~\cite{bib:deepcabac,bib:deepcabac2} is such a pioneer work on the model parameter compression, using  context-based adaptive binary arithmetic coding~(CABAC) originally developed for H.264 advanced video coding. 
DeepCABAC adaptively determines the coding parameters, e.g., the degree of quantization step-size, depending on the available bandwidth. 
However, the conventional model parameter transmission with digital source coding may experience some problems for wireless fading channels.  
First, the bitstream obtained from the digital encoder is highly vulnerable to communication errors~\cite{bib:survey}.
When the channel signal-to-noise ratio (SNR) falls below a certain threshold, channel errors in the bitstream cause the failure of the parameter restoration owing to the all-or-nothing behavior in entropy and channel codings.
A restoration failure will disrupt training the global and local model parameters at the aggregator and distributed learners, and thus it will cause limited performance.
Second, the quality of the reconstructed parameters will saturate even with channel quality improvement owing to unrecoverable quantization errors. 
To exchange high-quality parameters across the distributed terminals via rapid fading channels, adaptive rate control of source and channel coding is needed to perform in real-time.

\begin{figure*}[t]
  \begin{center}
   \includegraphics[scale=0.6]{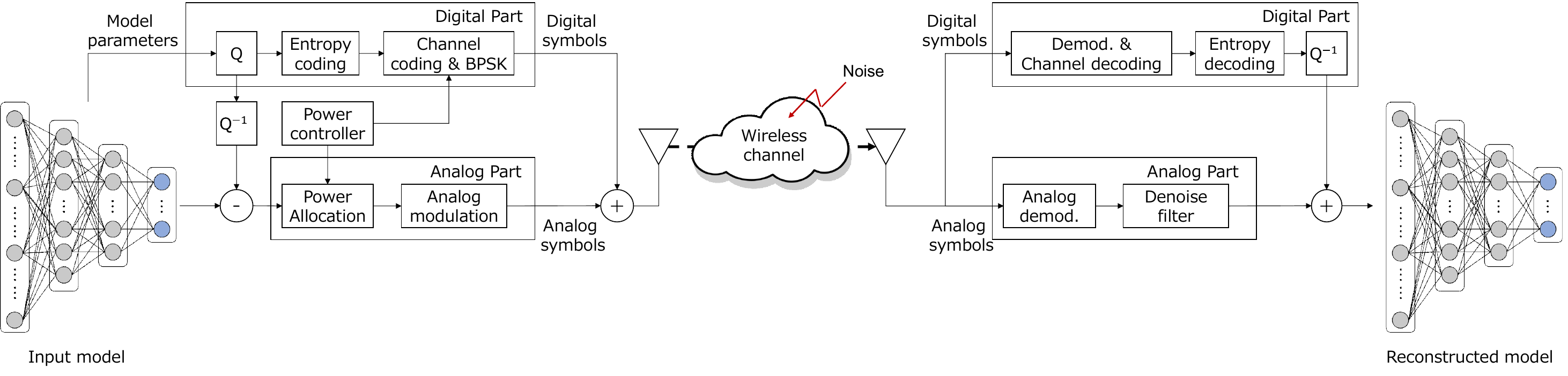}
   \caption{Sender side and receiver side operations in the proposed Federated AirNet, employing hybrid digital coding and analog coding to exchange model parameters of neural networks over wireless channels.}
   \label{fig:scheme}
  \end{center}
\end{figure*}

To overcome both issues of FL in wireless channels, analog-based solutions~\cite{bib:AirNet,bib:aircomp} have been proposed for the model parameter transmissions.
AirNet~\cite{bib:AirNet} is such a pioneer work on the analog-based transmission. 
AirNet directly maps the model parameters to the transmission symbols, i.e., analog modulation, and sends the analog-modulated symbols via wireless channels. 
This process avoids the issues mentioned above and the model restoration quality faithfully corresponds to the instantaneous channel condition. 
The analog over-the-air computation~(AirComp)~\cite{bib:aircomp,bib:aircomp2} is another solution to adopt the analog modulation for the model parameter transmissions in FL. 
All the learners simultaneously upload the analog-modulated parameters with channel inversion to the aggregator, and the aggregator can receive the aggregated model parameters from the superimposed waveforms in AirComp. 
However, the reconstruction quality of the linear mapping~(from source signals to channel signals) schemes greatly depends on the performance of the energy compaction technique for the source signals~\cite{bib:hybrid_theory}.

The above discussion points out three major challenges of wireless FL: 
1) to prevent the failure of the parameter restoration and leveling-off effect in neural network transmission;
2) to enhance neural network performance according to the wireless channel quality; 
and 3) to overcome the performance limitation owing to linear mapping. 
This paper proposes a novel hybrid digital-analog transmission of the model parameter transmissions for wireless FL systems, referred to as Federated AirNet, to overcome the three issues. 
The key idea of Federated AirNet is to integrate a low-rate digital-based solution into an analog-based solution, inspired by joint source-channel coding studies~\cite{bib:holoplus,bib:hybrid_first,bib:hda3}. 
The part of digital-based solution can compact the signal energy and prevent bit errors even with wireless channel fluctuations. 
The proposed Federated AirNet employs the analog modulation to supplementally send the residual signals between the original and digitally-encoded model parameters.
In consequence, the reconstruction quality of the model parameters can be gracefully improved according to the improvement of the wireless channel quality.

Focusing on an image classification task using ResNet-$18$, we demonstrate that the proposed Federated AirNet prevents the degradation of the classification accuracy owing to the channel quality fluctuation, outperforming the conventional digital-based and analog-based schemes. 
In addition, the proposed Federated AirNet can achieve higher classification accuracy with the limited number of model exchanges between the learners in the federation and the aggregator, leading to a lower demand of data traffic for the wireless FL systems.
The key contribution of this paper is three-fold as follows:
\begin{itemize}
\item To the best of our knowledge, the proposed Federated AirNet is the first scheme to realize a hybrid digital-analog transmission for model parameter transmissions of neural networks for wireless FL systems.
\item Since the signal energy can be reduced by the digital encoding, the proposed Federated AirNet can reconstruct more accurate model parameters as a function of the instantaneous wireless channel quality.
\item To deal with a precise symbol-level synchronization required in AirComp, we use a round-robin exchange in the Federated AirNet for quasi-asynchronous FL systems.
\item We demonstrate that our hybrid digital-analog FL scheme can outperform conventional digital- and analog-based schemes for the image classification task. 
\end{itemize}

\section{Hybrid Digital-Analog Federated AirNet}
\subsection{Architecture Overview}
Fig.~\ref{fig:scheme} shows the encoding and decoding architecture of our proposed Federated AirNet to exchange parameters of local model or global model between the aggregator and distributed leaners.  
Both encoder and decoder consist of digital and analog parts. 
The digital source encoder encodes the model parameters into a bitstream and takes channel coding with a low-rate convolutional code. 
The channel-coded bitstream is modulated by binary phase-shift-keying~(BPSK) format, i.e., mapped to the in-phase~(I) component.
After the digital encoding, the analog part obtains the residual parameters between its original and reconstructed models.
The residuals are then mapped to the quadrature~(Q) component to avoid interference to the BPSK-modulated symbols. 
For both digital-modulated and analog-modulated symbols, the power controller assigns unequal transmission power for error protection. 

On the receiver side, the digital-modulated and analog-modulated symbols are individually extracted from the received symbols. 
The digital-modulated symbols are demodulated and channel-decoded to obtain the decoded bitstream. 
The baseline of the model parameters can be reconstructed from the decoded bitstream using the digital decoder. 
The analog decoder denoises the analog-modulated symbols to obtain the residual parameters.
The receiver finally adds the reconstructed residuals to the digital-decoded output to obtain more accurate final output for updating its model parameters. 

\begin{figure}[t]
  \begin{center}
   \includegraphics[scale=0.35]{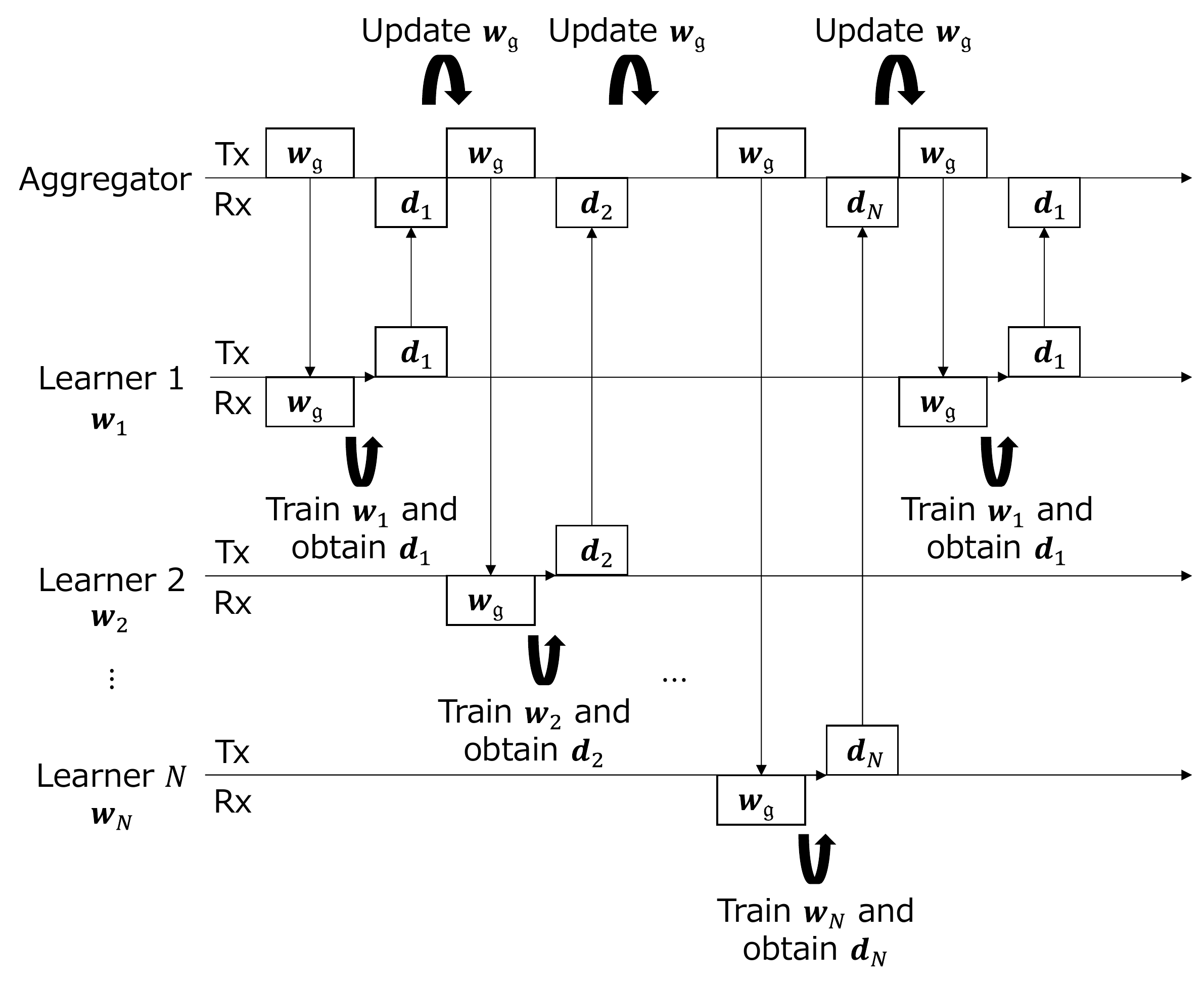}
   \caption{Timing diagram of the scheduling of local and global model transmissions.}
   \label{fig:timing}
  \end{center}
\end{figure}

\subsection{Scheduling of Local and Global Model Transmissions}
We consider the use of a round-robin scheduling across each learner in the federation and the aggregator to update local and global model parameters. 
Fig.~\ref{fig:timing} shows the timing diagram of the local and global model transmissions between the aggregator and $N$ learners.
Specifically, the aggregator sends the initial parameters of the global model to learner~$l$ using the above-mentioned hybrid digital-analog transmission. 
Given the received parameters of the global model $\bm{w}_\mathcal{G}$, the local model parameters for learner $l$ are initialized as $\bm{w}_l = \bm{w}_\mathcal{G}$. 
After updating the local model parameters using locally available dataset, the updated parameters $\hat{\bm{w}}_{l}$ are sent back to the aggregator in the difference form of $\bm{d}_{l} = \hat{\bm{w}}_{l} - \bm{w}_\mathcal{G}$ through the hybrid digital-analog transmission.

The aggregator then updates the global model parameters based on the received differences $\hat{\bm{d}}_{l}$ and pre-defined learning rate $l_r$. 
Specifically, the global model is updated as follows:
\begin{equation}
    \hat{\bm{w}}_\mathcal{G} = \bm{w}_\mathcal{G} + l_r \cdot \hat{\bm{d}}_{l},
\end{equation}
The updated global model parameters are sent to the next learner~$l+1$. 
Subsequently, the above procedures are repeated over $N$ learners up to a certain number of iterations.

\subsection{Energy Compaction}
The digital part of the Federated AirNet is based on DeepCABAC~\cite{bib:deepcabac} for encoding the model parameters into the bitstream. 
The digital encoder extracts the model parameters layer-by-layer in row-major order.
The quantization levels are defined based on the quantization step-size $\Delta$. 
The digital encoder then quantizes the parameters and compresses the quantized parameters using arithmetic coding with the lowest weighted rate distortion at a given bandwidth.  
Here, the quantization step-size is adapted according to the available bandwidth for the model parameter transmissions.

The analog part decodes the compressed bitstream to obtain the reconstructed baseline model parameters. 
The residuals between the original and reconstructed model parameters are analog-modulated. 
Specifically, given the original parameters $\bm{w}$ and reconstructed parameters from the digital part $\bar{\bm{w}}$, the analog part obtains the residual parameters as $\bm{r} = \bm{w} - \bar{\bm{w}}$, which are mapped to the Q component of constellation.

\subsection{Power Assignment for Digital and Analog Parts}
In prior to the transmissions, the Federated AirNet assigns unequal transmission power for the digital-modulated and analog-modulated symbols. 
The transmission power for the digital-modulated symbols should ensure the prevention of bit errors during transmissions. 
For this purpose, the power controller finds the required transmission power considering the required channel SNR, which guarantees the decoding bit-error rate~(BER) is below a target BER, as follows: 
\begin{equation}
P_\mathsf{th} = N_0 \cdot \gamma_0,
 \label{pth}
\end{equation}
where $\gamma_0$ is the required SNR, $P_\mathsf{th}$ is the required transmission power, and $N_0$ is the average noise power.
Given a total power budget of $P_\mathsf{t}$, the transmission powers for the digital-modulated symbol $P_\mathsf{d}$ and analog-modulated symbol $P_\mathsf{a}$ are determined as follows:
\begin{align}
P_\mathsf{d} &=
 \begin{cases}
    P_\mathsf{th}, & P_\mathsf{th} \leq P_\mathsf{t}, \\
    0, & \mathrm{otherwise},
 \end{cases} \qquad P_\mathsf{a} = P_\mathsf{t} - P_\mathsf{d}.
\end{align}

We let $x_{i}$ be the $i$th transmission symbol of the hybrid digital-analog transmission.
Each transmission symbol is a superposition of the digital-modulated symbol $x^{\langle\mathsf{d}\rangle}_{i}$ and
analog-modulated symbol $x^{\langle\mathsf{a}\rangle}_{i}$ as
follows:
\begin{equation}
x_{i} = x^{\langle\mathsf{d}\rangle}_{i}
 + \jmath\, x^{\langle\mathsf{a}\rangle}_{i},
\end{equation}
where $\jmath=\sqrt{-1}$ denotes the imaginary unit.
For error protection, both digital-modulated and analog-modulated symbols are scaled according to the assigned transmission power, i.e.,  $P_\mathsf{d}$ and
$P_\mathsf{a}$, respectively, as follows:
\begin{equation}
x^{\langle\mathsf{d}\rangle}_{i} = \sqrt{P_\mathsf{d}} \cdot b_{i},
 \qquad
x^{\langle\mathsf{a}\rangle}_{i} = m \cdot s_{i},
\end{equation}
where $b_{i}\in \mathbb{X} = \left\{\pm 1 \right\}$ is the
BPSK-modulated symbol and $r_{i} \in \bm{r}$ is the $i$th residual.
Here, $m$ is a normalization factor across the residual parameters as follows:
\begin{align}
\label{eq:water}
m=\sqrt{ \frac{M P_\mathsf{a}}{\sum_j \lambda_j}},
\end{align}
where ${\lambda}_{i}=|r_i|^2$ is the power of $i$th residual, and $M$ is the number of model parameters. 

The receiver obtains the symbols impaired by wireless links:
\begin{align}
y_{i} = x_{i} + n_{i},
\end{align}
where $y_{i}$ is the $i$th received symbol, and $n_{i}$ is an effective AWGN with a noise variance of $\sigma^2$.
In this model, the noise variance includes the effect of instantaneous fading gain.

\subsection{Decoding}
The receiver extracts both digital-modulated  and analog-modulated symbols from each received symbol's I and Q components, respectively. 
For digital-modulated symbols, the digital decoder calculates log-likelihood ratio~(LLR) values for all the received symbols~\cite{bib:fuji_cs} and feeds the LLR values into the Viterbi decoder to obtain a bitstream.
An arithmetic decoder finally decodes the bitstream to reconstruct the baseline of the transmitted parameters $\bar{\bm{w}}$.

For the analog-modulated symbol, i.e., $\Im(y_{i})$, the receiver denoises the received symbol to reconstruct the residuals:
\begin{align} 
\label{eq:mmse}
\hat{r}_{i} = \frac{m}{m^2 + \sigma^2} \cdot \Im(y_{i}).
\end{align}
We add the decoded residuals $\hat{\bm{r}}$ to the reconstructed baseline model parameters from the digital part $\bar{\bm{w}}$ to obtain the final output of the model parameters.

\section{Performance Evaluations}
\subsection{Experimental Settings}
\subsubsection{\textbf{Settings of Learners and Aggregator}}
We consider $N=10$ learners and one aggregator to share the same DNN model for FL systems. 
Here, each learner and the aggregator are connected via an wireless channel. 
We consider the number of transmissions between the aggregator and each learner is ten times. 
Each learner takes ten epochs for training before replying to the aggregator. 

\subsubsection{\textbf{Dataset}} 
We consider an image classification task using a benchmark CIFAR-10 dataset, which contains around $60{,}000$ color images of size $32 \times 32$ pixels labeled with ten different classes.  
We equally divide the training dataset of $50{,}000$ images into ten learners, and thus each learner possesses $5{,}000$ private images for training. 
We consider $4{,}000$ held-out images for evaluation of prediction performance. 

\subsubsection{\textbf{Network Model}}
All the parties, aggregator and learners, share the same neural networks of ResNet-$18$ having $11$ million of model parameters.
ResNet-$18$ consists of $18$ layers, including $17$ convolutional layers, an average pooling, and a fully connected layer. 
Ten outputs of DNN correspond to the ten classes of the CIFAR-10 dataset. 
We adopt the cross-entropy function as a loss function and momentum stochastic gradient descent with $0.9$ momentum, $5 \times 10^{-4}$ weight decay, and $0.1$ learning rate to update the model.
We use top-$1$ classification accuracy as a performance metric of DNN.

\subsubsection{\textbf{Comparative Schemes}}
We compare Federated AirNet to two state-of-the-art digital and analog approaches, specifically, DeepCABAC and AirNet. 
In the DeepCABAC scheme, the compressed bitstream is channel coded by a half-rate convolutional code with a constraint length of $8$ and digitally modulated by BPSK or $4$-QAM formats. 
The AirNet scheme directly maps each of the power-normalized model parameters onto a transmission symbol in analog modulation. 
The proposed Federated AirNet takes a half-rate convolutional code with a constraint length of $8$ and BPSK modulation format for the encoded bitstream obtained from DeepCABAC.
The BPSK-modulated symbols are superposed with the analog-modulated residuals.
We set $\gamma_0 = 5$~dB based on a preliminary evaluation.
Note that the receiver will be unable to update the model parameters in the case when arithmetic decoding fails to decompress the bitstream due to errors.

\begin{figure}[t]
  \centering
   \subfloat[Global model]{\includegraphics[width=8.8cm]{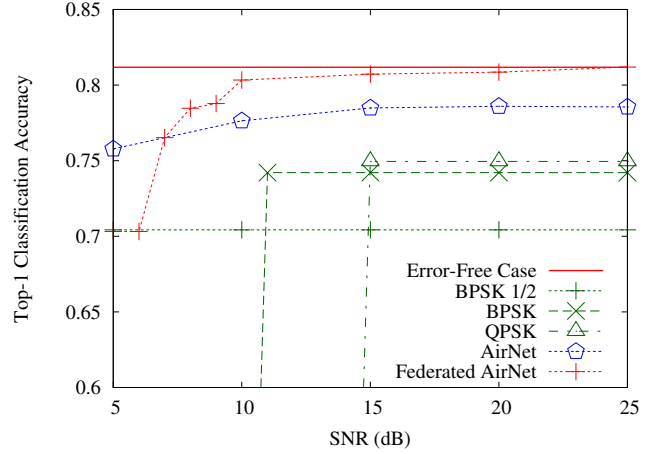}} \\
   \subfloat[Local model]{\includegraphics[width=8.8cm]{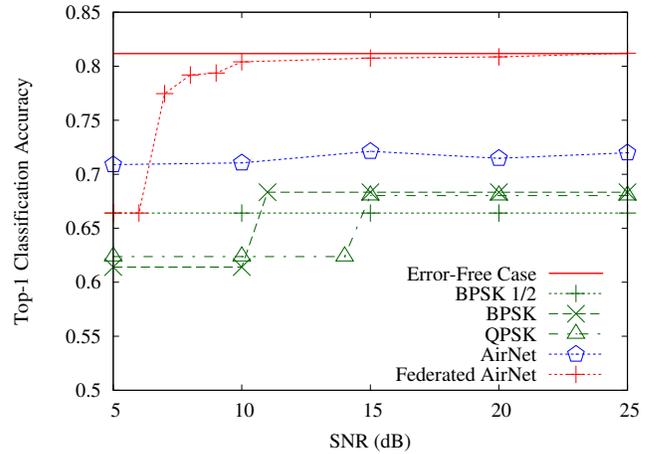}}
\caption{Average top-$1$ classification accuracy of global and local models as a function of wireless channel SNR after $10$-time transmissions. 
The available number of symbols transmitted over wireless channels is $6.0$~Msymbols.}
  \label{fig:snrvsacc}
\end{figure}

\begin{figure}[t]
  \centering
   \subfloat[Global model]{\includegraphics[width=8.8cm]{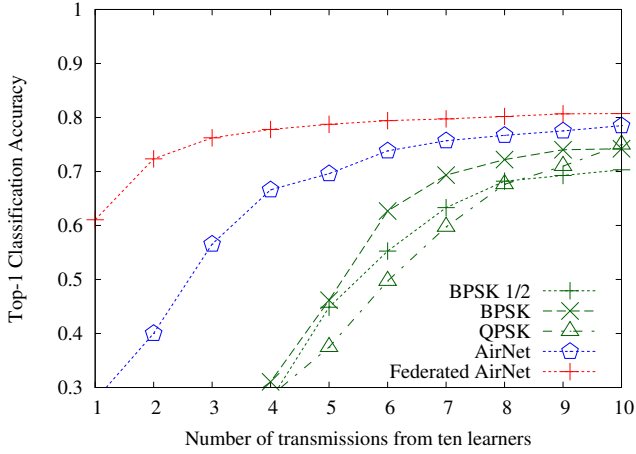}} \\
   \subfloat[Local model]{\includegraphics[width=8.8cm]{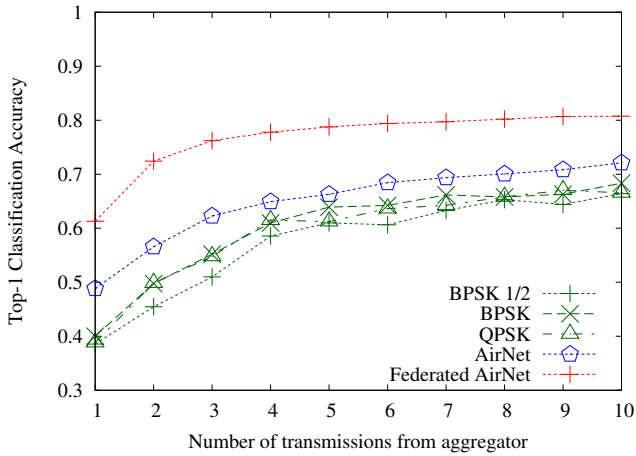}}
\caption{Average top-$1$ classification accuracy of global and local models as a function of the number of transmissions from aggregator/ten learners. 
The wireless channel SNR is $15$~dB and an available number of symbols transmitted over wireless channels is $6.0$~Msymbols.}
  \label{fig:uploadingvsacc}
\end{figure}

\begin{figure}[t]
  \centering
   \subfloat[Global model]{\includegraphics[width=8.8cm]{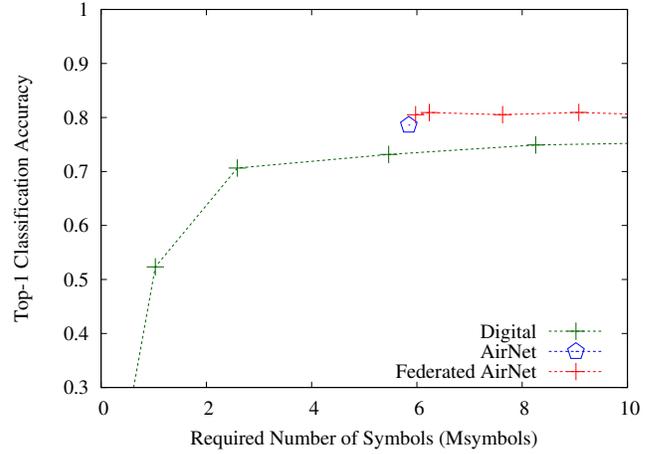}} \\
   \subfloat[Local model]{\includegraphics[width=8.8cm]{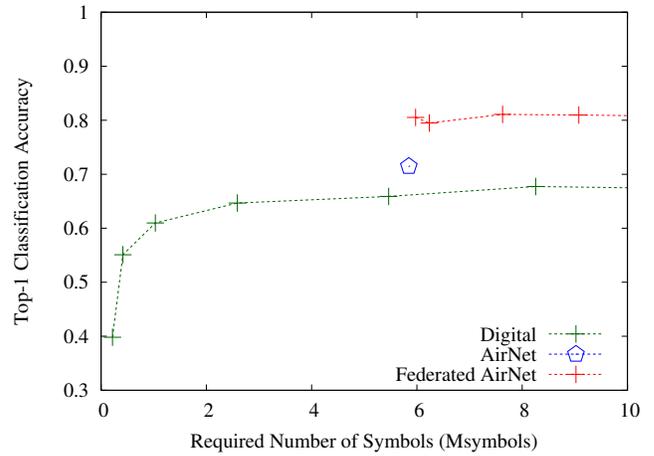}}
\caption{Average top-$1$ classification accuracy of global and local models as a function of the required number of symbols transmitted over wireless channels for a wireless channel SNR of $20$~dB.}
  \label{fig:ratevsacc}
\end{figure}

\subsection{Effect of Wireless Channel SNR}
We first evaluate the effect of the wireless channel quality on the classification accuracy. 
Figs.~\ref{fig:snrvsacc}~(a) and (b) show the average top-$1$ classification accuracy of global and local models, respectively, as a function of wireless channel SNR when an available number of transmission symbols is at most $6.0$~Msymbols. 
The number of transmission iterations from learners to aggregator is $10$. 
We can see the following results. 
\begin{itemize}
    \item In both local and global models, the proposed scheme yields the best accuracy, in higher channel SNR regimes, achieving near error-free performance.
    \item The digital-based schemes suffers from a sudden degradation of the classification accuracy in low SNR regimes. 
    \item The AirNet scheme prevents the drastic degradation in accuracy since it does not rely on quantization and entropy coding. However, the accuracy will be limited in high SNR regimes. 
    \item The proposed scheme has a lower classification accuracy than the analog-based scheme at the channel SNR below $6$~dB because the target threshold was set as $\gamma_0=5$~dB.
\end{itemize}

\subsection{Effect of Transmission Times}
The previous section showed that the proposed scheme yields better classification accuracy, especially in higher wireless channel SNR regimes for $10$-iteration FL exchanges.
This section discusses the effect of the number of transmissions between the aggregator and learners on classification accuracy. Figs.~\ref{fig:uploadingvsacc}~(a) and (b) show the average top-$1$ classification accuracy of global and local models, respectively, as a function of the transmission times at a channel SNR of $15$~dB.

The proposed Federated AirNet achieves the best accuracy irrespective of the number of transmissions. 
The classification accuracy of both global and local models exceeds $0.72$ after two transmission times.
Whereas, the AirNet and digital-based schemes require at least $5$ times and $7$ times transmissions, respectively, for global model learning. 
It suggests that the proposed Federated AirNet can reduce the total amount of traffic to achieve the same accuracy in the digital-based and analog-based schemes.

\subsection{Effect of Available Bandwidth}
We finally discuss the effect of the bandwidth requirements for the model parameter transmissions. 
Here, the digital-based and the proposed schemes control the quantization step-size $\Delta$ to evaluate the classification accuracy under different bandwidth requirements. 
The required bandwidth of an analog-based scheme depends on the number of the parameters, and thus AirNet and the analog part of the proposed Federated AirNet schemes use a fixed rate. 

Figs.~\ref{fig:ratevsacc}~(a) and (b) show the average top-$1$ classification accuracy of global and local models, respectively, as a function of the required number of symbols transmitted over wireless channels at a channel SNR of $20$~dB.
Even though the required transmission symbols increase, the proposed scheme keeps the best accuracy. 
On the other hand, it requires approximately $5.9$~Msymbols on average for model parameter transmissions. 
We note some studies in~\cite{bib:unequalblock,bib:SKmapping} discussed how to discard the analog-modulated symbols in bandwidth-constrained wireless channels. 
We will leave how to satisfy the narrow-band limitation in the proposed scheme as the future work. 

\section{Conclusion}
This paper proposed Federated AirNet to realize a novel hybrid digital-analog transmission of DNN model parameters for wireless federated learning. 
The main idea of the proposed scheme is to digitally send the baseline of the global and local model parameters whereas the residuals of the model parameters are sent in an analog manner. 
We show that integrating the digital and analog solutions can achieve good performance in terms of the image classification accuracy in both narrow-band and broadband scenarios.

\section*{Acknowledgment}
T. Fujihashi was partly supported by JSPS KAKENHI Grant Number 20K19783.

\bibliographystyle{IEEEtran}
\bibliography{./hybrid}

\end{document}